\documentstyle[prd,preprint,tighten,aps,eqsecnum,%
amssymb,newlfont]{revtex}
\begin{document}
\preprint{
\begin{tabular}{r}
UWThPh-1999-61\\
December 1999
\end{tabular}
}
\draft
\title{Two--Particle Correlations and Meson--Antimeson Mixing Effects}
\author{G.V. Dass}
\address{Physics Department, Indian Institute of Technology\\
Powai, Bombay - 400\,076, India}
\author{W. Grimus}
\address{Institut f\"ur Theoretische Physik,
Universit\"at Wien\\
Boltzmanngasse 5,
A-1090 Vienna, Austria}
\maketitle
\begin{abstract}
We discuss 2-particle correlations which arise in the time evolution of C-odd
and C-even meson--antimeson states of flavoured neutral mesons. 
In order to keep our discussion
general, we do not use the Weisskopf -- Wigner
approximation. Possible deviations from quantum-mechanical coherence
effects are parameterized by a so-called decoherence parameter $\zeta$. In
particular, we study the $\zeta$-dependence of the 
asymmetry of unlike and like-flavoured events which was recently observed 
experimentally in the $K^0 \bar{K}^0$ system. In this $\zeta$-dependence, we 
point out some important general features which
do not rely on the Weisskopf -- Wigner approximation. Some other
related results are derived more generally than in the literature.
\end{abstract}


\section{Introduction}

Quantum-mechanical (QM) correlations arising in the time evolution of
the 2-particle wave functions 
$| f \otimes \bar f - \bar f \otimes f \rangle$,
where $f$ is some flavoured neutral meson like $f = K^0, B^0_d$ and 
$\bar f$ is its antiparticle, have been studied recently (see
Refs.~\cite{BGH99,BG97,dass,BG98} and papers cited therein). The aim
in these papers was to use data in order to see whether the
correlations expected on the basis of quantum mechanics were present in full,
partial or zero strength. To achieve this, 
a so-called ``decoherence parameter''
$\zeta$ was introduced \cite{eberhard}. 
The conclusions have been in favour of ``full
strength'', as far as the present data can say. 
The usual phenomenology of the 
$(f, \bar f)$ complex has been utilized in the various analyses. This
phenomenology is based on the Weisskopf--Wigner approximation
(WWA) \cite{WWA}. 

Because of the importance of the conclusions,
it is advisable to make the analysis as model-independent as possible.
The purpose of the present note is to see how far one may go
without involving the WWA. By the same token, the numerical predictive
power is, of course, greatly reduced if the WWA is not invoked. 
The WWA is considered in this
paper only for comparison with the general results which we derive. We
shall make some corresponding remarks on the time evolution of
the state $| f \otimes \bar f + \bar f \otimes f \rangle$ also.
In addition, some related results will be derived more generally
than in the literature.

The plan of the paper is as follows. In Sec.~II, we give the
basic framework required for discussing the time evolution of
the two-particle states 
$| f \otimes \bar f \pm \bar f \otimes f \rangle$
in the general and various special cases. Sec.~III is devoted to
the evaluation of the recently observed experimental asymmetry
\cite{CPLEAR98} (see also the relevant experimental papers cited
in Refs.~\cite{BG97,dass}) between the rates of the production
of like-flavoured 
(viz. $| f \otimes f \rangle$ and $| \bar f \otimes \bar f \rangle$)
and unlike-flavoured
(viz. $| f \otimes \bar f \rangle$ and $| \bar f \otimes f \rangle$)
states of the $(f, \bar f)$ complex. In Sec.~IV, we note that
some purely QM results hold more generally than the derivations
in the literature; this includes also cases within the WWA. Sec.~V
is devoted to a systematic discussion of the basis needed for
a vanishing of the experimental asymmetry under the Furry
hypothesis (absence of QM interference terms, corresponding to
$\zeta = 1$) \cite{furry}; we show that the conclusion of
Ref.~\cite{BGH99} needs generalization. In Sec.~VI, we try to
see how the introduction of $\zeta$ effectively induces a CP
violation; we also study a special configuration considered in
Ref.~\cite{BGH99}. Our derivation is more general and much
simpler than in the literature.

\section{The formalism}

Starting from the states $| f \rangle$ and $| \bar f \rangle$
initially, i.e., at proper time $t=0$, the time evolution is given by
\cite{khalfin,dass99} 
\begin{equation}
\begin{array}{lll}
| f \rangle & \stackrel{t}{\rightarrow} &
a(t) | f \rangle + b(t) | \bar f \rangle +
\sum_i e_i(t) | \rho_i \rangle \,, \\[1mm]
| \bar f \rangle & \stackrel{t}{\rightarrow} &
\bar b(t) | f \rangle + \bar a(t) | \bar f \rangle +
\sum_i \bar e_i(t) | \rho_i \rangle \,,
\end{array}
\end{equation}
where the $| \rho_i \rangle$ ($i=1,2,\ldots$) are other states,
orthogonal to $| f \rangle$ and $| \bar f \rangle$, produced with
coefficients $e_i$ and $\bar e_i$. We shall need only the amplitudes 
$a$, $\bar a$, $b$ and $\bar b$. While $a$ and $\bar a$ denote the
diagonal transitions 
$| f \rangle \to | f \rangle$ and $| \bar f \rangle \to | \bar f
\rangle$, respectively, $b$ and $\bar b$ denote the ``mixing'' or
non-diagonal transitions $| f \rangle \to | \bar f \rangle$ and 
$| \bar f \rangle \to | f \rangle$, respectively. We shall keep the
amplitudes $a$, $\bar a$, $b$ and $\bar b$ unknown and independent,
unless otherwise stated.

The WWA provides a model for these four amplitudes. Note
that by definition \cite{dass99} we have
\begin{equation}\label{initial}
a(0) = \bar a(0) = 1 \,, \quad
b(0) = \bar b(0) = 0 \,, \quad
e_i(0) = \bar e_i(0) = 0 \quad \forall i \,.
\end{equation}
Taking $| \bar f \rangle = \mathcal{CP} | f \rangle$, CP violation
means that $a \neq \bar a$ and/or $b \neq \bar b$. It is useful to
introduce the CP-odd amplitudes
\begin{equation}\label{vV}
v = \frac{1}{2} (a - \bar a), \quad V = \frac{1}{2} (b - \bar b)
\end{equation}
and the CP-even amplitudes
\begin{equation}\label{AB}
A = \frac{1}{2} (a + \bar a), \quad B = \frac{1}{2} (b + \bar b) \,.
\end{equation}
Then, $v=0$ corresponds to CPT (and CP) invariance, and $V=0$ to CP
(and T) invariance.

The $(f, \bar f)$ complex may be equally well 
described in terms of the normalized superpositions 
\begin{equation}\label{f1f2}
  \begin{array}{ll}
| f_1 \rangle = p_1 | f \rangle + q_1 | \bar f \rangle \,, &
|p_1|^2 + |q_1|^2 = 1 \,, \\[1mm]
| f_2 \rangle = p_2 | f \rangle - q_2 | \bar f \rangle \,, &
|p_2|^2 + |q_2|^2 = 1 \,,
  \end{array}
\end{equation}
where $p_{1,2}$ and $q_{1,2}$ are some complex constants. Exploiting the
freedom of the three unmeasurable phases, one may use the Eberhard phase
convention \cite{phase} to write
\begin{equation}
  \begin{array}{ll}
p_1 = e^{i \theta/2} \cos \alpha_1 \,, & 
q_1 = e^{-i \theta/2} \sin \alpha_1 \,, \\[1mm]
p_2 = e^{-i \theta/2} \cos \alpha_2 \,, & 
q_2 = e^{i \theta/2} \sin \alpha_2 \,.
  \end{array}
\end{equation}
Note that the ranges of these parameters can be confined to 
$0 \leq \alpha_{1,2} \leq \pi/2$ and 
$-\pi/2 \leq \theta \leq \pi/2$ without loss
of generality. Instead of $\alpha_1$ and $\alpha_2$, one can also use the 
angles $\sigma$, $\delta$ defined as 
\begin{equation}
\sigma = \frac{\pi}{2} - (\alpha_1 + \alpha_2) \,, \quad
\delta = \alpha_1 - \alpha_2 \,.
\end{equation}
Thus the three significant parameters occurring in Eq.(\ref{f1f2})
can be taken to be the real
CP-violating parameters $\theta$, $\sigma$ and $\delta$. The scalar product
of the states in Eq.(\ref{f1f2}) can be expressed as
\begin{equation}
\langle f_2 | f_1 \rangle = 
\sin \sigma \cos \theta + i \cos \delta \sin \theta \,.
\end{equation}
As a corollary, the CP eigenstates
\begin{equation}\label{f+-}
| f_\pm \rangle = \frac{1}{\sqrt{2}} 
\left( | f \rangle \pm | \bar f \rangle \right) 
\end{equation}
arise with the choice 
\begin{equation}
\theta = \sigma = \delta = 0 \,, 
\end{equation}
corresponding to
\begin{equation}\label{CPcons}
\alpha_{1,2} = \frac{\pi}{4} \quad \mbox{and} \quad
p_{1,2} = q_{1,2} = \frac{1}{\sqrt{2}} \,.
\end{equation}

The WWA is characterized by the introduction of independently propagating
states $| f_{S,L} \rangle$ which are superpositions (with constant complex
coefficients $p_{S,L}$ and $q_{S,L}$) of $| f \rangle$ and $| \bar f \rangle$:
\begin{equation}\label{fSfL}
  \begin{array}{ll}
| f_S \rangle = p_S | f \rangle + q_S | \bar f \rangle \,, &
|p_S|^2 + |q_S|^2 = 1 \,, \\[1mm]
| f_L \rangle = p_L | f \rangle - q_L | \bar f \rangle \,, &
|p_L|^2 + |q_L|^2 = 1 \,.
  \end{array}
\end{equation}
These states have the time evolution
\begin{equation}
| f_{S,L} \rangle \stackrel{t}{\rightarrow} \Theta_{S,L}(t)\, | f_{S,L} \rangle
\end{equation}
with
\begin{equation}
\Theta_{S,L}(t) = \exp 
\left\{ -i t \left( m_{S,L} - \frac{i}{2} \Gamma_{S,L} \right) \right\} \,,
\end{equation}
where $m_{S,L}$ and $\Gamma_{S,L}$ are the usual masses and widths for 
$| f_{S,L} \rangle$. There are only two definite $t$-dependences now, given by 
$\Theta_{S,L}(t)$. The initial conditions (\ref{initial}) imply that $v$, $V$
and $B$ must have the $t$-dependence $[\Theta_S(t)-\Theta_L(t)]$. Indeed, one
finds (see Ref.~\cite{kabir} for a review)
\begin{eqnarray}
a + \bar a & = & \Theta_S + \Theta_L \,, 
\label{a+} \\
a - \bar a & = & \left( \frac{p_S}{q_S} - \frac{p_L}{q_L} \right) b \,, 
\label{a-} \\
\bar b     & = & \left( \frac{p_S p_L}{q_S q_L} \right) b \,, 
\label{bbar} \\
b          & = & \frac{q_S q_L}{ p_S q_L + p_L q_S } (\Theta_S - \Theta_L) \,.
\label{b}
\end{eqnarray}
Also, Eqs.(\ref{a-}) and (\ref{bbar}) imply the relations
\begin{equation}\label{vvwwa}
v(t) = c_1 B(t) \quad \mbox{and} \quad V(t) = c_2 B(t)
\end{equation}
with constants $c_{1,2}$ being given by the relations (\ref{a-}) and
(\ref{bbar}). The two relations in Eq.(\ref{vvwwa}) can be conceived as
``general'' consequences of the WWA, while the stronger relations (\ref{a+})
and (\ref{b}) are necessary for explicit calculations using the WWA fully.
In Eq.(\ref{vvwwa}) we note that
\begin{eqnarray}
\label{c1}
\mbox{CPT (CP) invariance} & \Rightarrow & c_1 = 0 \,, \\
\label{c2}
\mbox{CP (T) invariance} & \Rightarrow & c_2 = 0 \,.
\end{eqnarray}
Note that the $c_{1,2}$ are defined in the WWA, but not in general.

The above 1-particle description is required in the analysis of the time
evolution of the 2-particle states
\begin{equation}\label{state}
| f \otimes \bar f + \epsilon \bar f \otimes f \rangle \,,
\end{equation}
where $\epsilon = +1$ and $-1$ denote the C-even and C-odd states,
respectively. Examples of $\epsilon = -1$ are the $\phi$ meson, the 
$\Upsilon(4\mathrm{S})$ and the $^3\mathrm{S}_1$ state
of $p \bar p$ annihilation
at rest \cite{CPLEAR98}. An example of $\epsilon = +1$ would be the state 
$| s \rangle$ obtained by radiative decays of $\Upsilon(4\mathrm{S})$:
\begin{equation}
\Upsilon(4\mathrm{S}) \to s + \gamma \,.
\end{equation}
Denoting the proper times involved in the time evolution of the states
(\ref{state}) by $t'$ and $t$ and using the notation
\begin{equation}\label{notation}
a' \equiv a(t'), \quad a \equiv a(t), \quad \ldots ,
\end{equation}
we can write the time evolution in terms of these eight amplitudes which, in
general, are all unknown \cite{dass99}.

Different bases $(f_1,f_2)$ (\ref{f1f2}) 
used in expressing the $t'$, $t$-evolution of
(\ref{state}) correspond to different $p_{1,2}$ and $q_{1,2}$. One can
introduce the $t$-dependent amplitudes $A_{1,2}$ and $\bar A_{1,2}$ as 
\begin{equation}\label{timeevol}
| f_{1,2} \rangle \stackrel{t}{\rightarrow} 
A_{1,2}\, | f \rangle + \bar A_{1,2}\, | \bar f \rangle
\end{equation}
with
\begin{equation}\label{AA}
\left( \begin{array}{c} A_1 \\ A_2 \end{array} \right) =
\left( \begin{array}{lr} p_1 & q_1 \\ p_2 & -q_2 \end{array} \right)
\left( \begin{array}{c} a \\ \bar b \end{array} \right) \,, \quad
\left( \begin{array}{c} \bar A_1 \\ \bar A_2 \end{array} \right) =
\left( \begin{array}{lr} p_1 & q_1 \\ p_2 & -q_2 \end{array} \right)
\left( \begin{array}{c} b \\ \bar a \end{array} \right) \,.
\end{equation}
The choice $p_1 = -q_2 = 1$, $q_1 = p_2 = 0$ would correspond to the 
$(f, \bar f)$ basis. The notation (\ref{notation}) indicating the association
of the primed amplitudes with $t'$ and of the unprimed amplitudes with $t$
will apply also to $A_{1,2}$ and $\bar A_{1,2}$. Possible CP invariance of
amplitudes refers to $a$, $\bar a$, $b$ and $\bar b$, viz. $v = 0$ and/or 
$V = 0$. On the other hand, the general mixing parameters lead to the basis
provided by the CP eigenstates $| f_\pm \rangle$ of Eq.(\ref{f+-}), 
if the choice
(\ref{CPcons}) is made. If, however, one specializes to the WWA, the
amplitudes and the mixing parameters ($p_{S,L}$; $q_{S,L}$) are
intimately related; the CP property of the amplitudes is given by the
choice (\ref{c1}) and (\ref{c2}) of these mixing parameters.

Special cases of the completely general amplitudes could be
(i)
the general CP-invariant amplitudes ($v=V=0$) without the WWA,
(ii)
the WWA amplitudes, which may also obey CP invariance ($c_1=c_2=0$).
Similarly, the special cases of the general basis states 
$| f_{1,2} \rangle$ could be the $| f_\pm \rangle$, or the flavour
states $| f \rangle$, $| \bar f \rangle$, or the WWA states 
$| f_{S,L} \rangle$  characterized by independent propagation. Of
course, one can also have $| f_{S,L} \rangle$ becoming 
$| f_\pm \rangle$ as a further assumption.

\section{Evaluation of the experimentally observed asymmetry 
in the general and special cases} 
\label{evaluation}

The experimental observable recently investigated 
\cite{BGH99,BG97,dass,BG98,CPLEAR98} utilizes the production of the
like-flavoured states 
$| f \otimes f \rangle$ and $| \bar f \otimes \bar f \rangle$ 
and the unlike-flavoured states
$| f \otimes \bar f \rangle$ and $| \bar f \otimes f \rangle$. We use $t'$ for
the time of the measurement of the first meson and $t$ for the second
meson. The CP-even observable \cite{CPLEAR98} is the asymmetry 
$\mathcal{A}(t',t)$ between the numbers of all like-flavoured (viz. $f f$ and
$\bar f \bar f$) events and all unlike-flavoured 
(viz. $f \bar f$ and $\bar f f$) events:
\begin{equation}\label{asymm}
\mathcal{A}(t',t) = 
\frac{\mathcal{N}_\mathrm{unlike}(t',t) - \mathcal{N}_\mathrm{like}(t',t)}%
     {\mathcal{N}_\mathrm{unlike}(t',t) + \mathcal{N}_\mathrm{like}(t',t)}\,.
\end{equation}

We first concentrate on the state provided by the choice 
$\epsilon = -1$ in Eq.(\ref{state}). This CP-odd state 
can be rewritten in terms of the general basis
(\ref{f1f2}). Taking into account the time evolution, one simply gets
\begin{equation}\label{+}
| f_1(t') \otimes f_2(t) - f_2(t') \otimes f_1(t) \rangle \,,
\end{equation}
where, for the channels relevant to (\ref{asymm}), 
\begin{equation}
| f_{1,2}(t) \rangle = 
A_{1,2}(t) | f \rangle + \bar{A}_{1,2}(t) | \bar f \rangle \,,
\end{equation}
and, similarly, for the argument $t'$.
Thus, for the calculation of $\mathcal{A}(t',t)$ we use Eq.(\ref{timeevol}) 
for the time
evolution and multiply the interference term by the factor $(1-\zeta)$
in order to parameterize deviations from quantum mechanics 
\cite{eberhard,BGH99,BG97,dass,BG98}. This leads to the general
expression
\begin{equation}\label{A}
\mathcal{A}(t',t) = \frac{N_1(t',t) + 2(1-\zeta) D_1(t',t)}%
                   {N_2(t',t) + 2(1-\zeta) D_2(t',t)}  \,,
\end{equation}
where
\begin{eqnarray}
N_{1,2} & = &
\left[ |A'_2|^2 \mp |\bar A'_2|^2 \right] 
\left[ |\bar A_1|^2 \mp |A_1|^2 \right] 
+ ( t \leftrightarrow t' ) \,, \label{N} \\
D_{1,2} & = & \mbox{Re} \left\{ 
\left[-A'^*_2 A'_1 \pm \bar A'^*_2 \bar A'_1 \right]
\left[ \bar A^*_1 \bar A_2 \mp A^*_1 A_2 \right] \right\} \,.
\label{D}
\end{eqnarray}
Here, the real decoherence parameter $\zeta$ \cite{eberhard} is just
one way of expressing possible departures from quantum
mechanics. While quantum mechanics
corresponds to $\zeta = 0$, complete decoherence (Furry's hypothesis
\cite{furry}) corresponds to $\zeta = 1$, which means absence of the QM
correlations expressed by the interference terms $D_{1,2}$ of Eq.(\ref{D}). 

In order to be able to consider the case of small departures 
from CP conservation (i.e., the basis states $| f_{1,2} \rangle$ 
are close to $| f_\pm \rangle$ and the CP-violating amplitudes $v$ and 
$V$ of Eq.(\ref{vV}) are small relative to $A$ and $B$ of Eq.(\ref{AB})),
we write the amplitudes $A_{1,2}$ and $\bar A_{1,2}$, up to first order in CP
violation, as
\begin{equation}\label{Adelta}
\begin{array}{rcr}
\renewcommand{\arraystretch}{1.5}
     A_1 & = &  \frac{1}{\sqrt{2}} (A+B+\Delta_1) \,, \\
\bar A_1 & = &  \frac{1}{\sqrt{2}} (A+B-\Delta_1) \,, \\
     A_2 & = &  \frac{1}{\sqrt{2}} (A-B+\Delta_2) \,, \\
\bar A_2 & = & -\frac{1}{\sqrt{2}} (A-B-\Delta_2) \,,
\end{array}
\end{equation}
with
\begin{equation}\label{delta}
\Delta_1 = v-V + \beta_1 (A-B) \,, \quad 
\Delta_2 = v+V + \beta_2 (A+B) \,.
\end{equation}
The quantities 
\begin{equation}\label{pqdelta}
\begin{array}{l}
\frac{1}{\sqrt{2}}\, \beta_1 = p_1 - \frac{1}{\sqrt{2}} = 
\frac{1}{\sqrt{2}} - q_1 = 
\frac{1}{2\sqrt{2}}\, (\sigma + i\theta - \delta ) \,, \\[1mm]
\frac{1}{\sqrt{2}}\, \beta_2 = p_2 - \frac{1}{\sqrt{2}} = 
\frac{1}{\sqrt{2}} - q_2 =
\frac{1}{2\sqrt{2}}\, (\sigma - i\theta + \delta )
\end{array}
\end{equation}
denote the first order CP-violating corrections for
the coefficients $p_{1,2}$, $q_{1,2}$
of Eq.(\ref{f1f2}). Now it is easy to show that
$N_1$ and $D_2$ are quantities of second order of smallness and that $N_2$ and
$D_1$ have no contribution of first order of CP violation. Thus, neglecting
effects of second order of CP violation, the asymmetry (\ref{A}) becomes
\begin{equation}\label{A+}
\mathcal{A}(t',t)|_{\epsilon = -1} = (1-\zeta)
\mathcal{A}^\mathrm{QM}_{\epsilon = -1} \,,
\end{equation}
where
\begin{equation}\label{AQM}
\mathcal{A}^\mathrm{QM}_{\epsilon = -1} = 
\frac{2\, \mbox{Re}\, (A'^*_1 A^*_2 A'_2 A_1)}{|A'_1 A_2|^2 + |A'_2 A_1|^2}
\end{equation}
is the QM asymmetry neglecting second order of CP violation. Note that,
for Eq.(\ref{AQM}), we effectively have
\begin{equation}\label{CP0}
A_1 =  \bar A_1 = \frac{1}{\sqrt{2}} (A+B) \,, \quad 
A_2 = -\bar A_2 = \frac{1}{\sqrt{2}} (A-B) \,.
\end{equation}
The proportionality of $\mathcal{A}(t',t)$ to
$(1-\zeta)$, apart from terms of second order in CP violation, has
been noted in Ref.~\cite{BGH99} in explicit calculations
using the WWA fully and assuming CP conservation in mixing. 
Our derivation shows that the result does not depend on the WWA. 

Coming now to the case $\epsilon = +1$ and rewriting the state
(\ref{state}) in terms of the general basis (\ref{f1f2}), we obtain
\begin{equation}\label{state+}
| 2 p_2 q_2 f_1 \otimes f_1 - 2 p_1 q_1 f_2 \otimes f_2 +
( p_2 q_1 - p_1 q_2 ) ( f_1 \otimes f_2 + f_2 \otimes f_1 ) \rangle \,,
\end{equation}
where we have neglected an irrelevant normalization factor. If we
proceed now by replacing the states $| f_{1,2} \rangle$ by the
time-evolved states as in Eq.(\ref{+}), we see that there are several
 possibilities to introduce decoherence, in contrast to
Eq.(\ref{+}). However, there is
a drastic simplification if we assume CP invariance:
$p_1 q_1 = p_2 q_2 = 1/2$ and $p_2 q_1 - p_1 q_2 = 0$.
Then, if we assume CP invariance,
our starting point for $\epsilon = +1$ is given by
\begin{equation}\label{-}
| f_+(t') \otimes f_+(t) - f_-(t') \otimes f_-(t) \rangle \,,
\end{equation}
analogous to the state (\ref{+}). In analogy to Eq.(\ref{A+}), we obtain
\begin{equation}\label{A-}
\mathcal{A}(t',t)|_{\epsilon = +1} = (1-\zeta)
\mathcal{A}^\mathrm{QM}_{\epsilon = +1} \,,
\end{equation}
where
\begin{equation}\label{AQM-}
\mathcal{A}^\mathrm{QM}_{\epsilon = +1} = 
\frac{2\, \mbox{Re}\, (A'^*_1 A^*_1 A'_2 A_2)}{|A'_1 A_1|^2 + |A'_2 A_2|^2}
\,.
\end{equation}
Thus, in the case of $\epsilon = +1$ also, 
the asymmetry (\ref{asymm}) is proportional to
$(1-\zeta)$ if CP invariance holds; the amplitudes in Eq.(\ref{AQM-}) are
again given by Eq.(\ref{CP0}).

If the basis states are $| f \rangle$ and $| \bar f \rangle$, the
introduction of the decoherence parameter proceeds in the same way for
$\epsilon = \pm 1$, but the dependence of the asymmetry (\ref{asymm})
on $\zeta$ is not so simple:
\begin{equation}\label{Aff}
\mathcal{A}(t',t)|_{\epsilon = \pm1} =
\frac{\left( |a'|^2-|b'|^2 \right) \left( |a|^2-|b|^2 \right) +
      4\epsilon (1-\zeta)\, \mbox{Im}\, (a'^* b')\, \mbox{Im}\,(b^* a)}%
     {\left( |a'|^2+|b'|^2 \right) \left( |a|^2+|b|^2 \right) +
      4\epsilon (1-\zeta)\, \mbox{Re}\, (a'^* b')\, \mbox{Re}\,(b^* a)}\,,
\end{equation}
wherein the CP-violating amplitudes $v$, $V$ have been
dropped. This $\zeta$-dependence and the non-zero value of the
asymmetry for $\zeta = 1$ have been noted in Ref.~\cite{BGH99} 
(see also Refs.~\cite{BG97,dass,BG98}) for 
$\epsilon = -1$, in explicit calculations using the WWA fully. Again,
our derivation shows the result to be more general. In the case of the
Furry hypothesis ($\zeta = 1$) the non-zero asymmetry for $\epsilon =
+1$ is the same as for $\epsilon = -1$, as seen from
Eq.(\ref{Aff}). This is similar to the corresponding comparison
between the asymmetries (both zero) using the basis states 
$| f_\pm \rangle$ (see Eqs.(\ref{A+}) and (\ref{A-})).

\section{Purely quantum-mechanical results}
\label{purely}

It is easy to see that the basis independence (viz. $p_{1,2}$ and
$q_{1,2}$ can be arbitrary) of the net QM amplitude, starting from
Eq.(\ref{state}), for \emph{any} detected final states (which
may be made up of states other than $| f \rangle$ and
$| \bar f \rangle$ used in Section \ref{evaluation})
holds quite generally: (i) for $\epsilon = +1$ and $\epsilon = -1$,
(ii) arbitrary transition amplitudes from $| f \rangle$ and
$| \bar f \rangle$ to the detected final states. It is really a matter
of transferring (\ref{state}) to the $| f_{1,2} \rangle$ basis using
the transformation inverse to that in Eq.(\ref{f1f2}).  Then, upon
introducing transition amplitudes, one applies the 
transformations analogous to Eq.(\ref{AA}) for the chosen final
detected states. In the process, the $p_{1,2}$ and $q_{1,2}$ drop out
and the overall transition amplitude is seen to be the same as in the
$(f, \bar f)$ basis, in general. This basis independence
\cite{BGH99,BG97,dass,BG98} has been noted for $\epsilon = -1$ and
using the WWA.

As a corollary, one may note for $\zeta \neq 0$ (i.e., in the case of
departure from quantum mechanics), the value of $\zeta$ inferred from experimental
data is, in general, basis-dependent. Because the QM amplitude is
basis-independent, the corresponding probability $P_\mathrm{QM}$ is
basis-independent as well. Writing
\begin{equation}\label{PP}
P_\mathrm{QM} = P_\mathrm{Furry} + P_\mathrm{int} \,,
\end{equation}
both $P_\mathrm{Furry}$ and $P_\mathrm{int}$ are, in general,
basis-dependent (see, e.g., Eqs.(\ref{A}--\ref{D})). 
Comparing the $\zeta$-modified version
\begin{equation}\label{PPz}
P_\mathrm{QM} = P_\mathrm{Furry} + (1-\zeta) P_\mathrm{int} 
\end{equation}
of Eq.(\ref{PP}) with experimental data will lead  to different
numerical values for $\zeta$ for different bases because
$P_\mathrm{int}$ is basis-dependent. This has been noted in explicit
calculations with the WWA in Refs.~\cite{BGH99,BG97,dass,BG98}.

We now make some comments on QM calculations within the WWA, using
Eq.(\ref{vvwwa}) but not
Eqs.(\ref{a+}, \ref{b}). Within this general WWA one gets
\cite{khalfin,dass99,dass92} 
\begin{equation}\label{ppaa}
\frac{\mathcal{N}(f,f)}{\mathcal{N}(\bar f, \bar f)} =
\left( \frac{1-c_2}{1+c_2} \right)^2
\end{equation}
for $\epsilon = -1$. In this equation, 
$\mathcal{N}(\stackrel{\scriptscriptstyle (-)}{f},
\stackrel{\scriptscriptstyle (-)}{f})$ denotes the
number of events yielding 
$| \stackrel{\scriptscriptstyle (-)}{f} \otimes
\stackrel{\scriptscriptstyle (-)}{f} \rangle$. Thus, the two ``like''
states are produced in a proportion which is independent of $t'$ and
$t$, quite generally, within the WWA. The corresponding case of 
$\epsilon = +1$ requires the additional assumption of CPT invariance
($c_1 =0$), again with the result (\ref{ppaa}). If one requires the
corresponding ratio for the ``unlike'' states to be constant for
$\epsilon = \pm1$, one
obtains the constant to be unity for all $t'$ and $t$, if CPT
invariance ($c_1 =0$) holds:
\begin{equation}\label{paap}
\frac{\mathcal{N}(f, \bar f)}{\mathcal{N}(\bar f, f)} = 1 \,.
\end{equation}
In all the above cases $c_2 \neq 0$ is allowed. 
The results (\ref{ppaa}) and (\ref{paap}) for $\epsilon = \pm 1$ are
to be compared with the simple results
\begin{equation}\label{1}
\frac{\mathcal{N}(f,f)}{\mathcal{N}(\bar f, \bar f)} =
\frac{\mathcal{N}(f, \bar f)}{\mathcal{N}(\bar f, f)} = 1 
\end{equation}
at any $t'$, $t$, if CP invariance of the amplitudes holds ($v=V=0$),
even without the WWA. Eq.(\ref{1}) arises because the initial state is
a CP eigenstate, and under CP one has
\begin{equation}
| f \otimes f \rangle \to | \bar f \otimes \bar f \rangle
\quad \mbox{and} \quad
| f \otimes \bar f \rangle \to | \bar f \otimes f \rangle \,.
\end{equation}
While Eqs.(\ref{ppaa}) and (\ref{paap}) are based on the WWA
\emph{without} $V=0$ (in the form of $c_2 = 0$), Eq.(\ref{1}) does not
use the WWA, but it does use $V=0$. Eq.(\ref{ppaa}) does not need CPT
invariance (viz. $c_1 = 0$) for $\epsilon = -1$, but it does need 
$c_1 = 0$ for $\epsilon = +1$. The relation $c_1 = 0$ is also needed
for Eq.(\ref{paap}). Some of the above results have been noted in
explicit calculations using the WWA fully, namely Eqs.(\ref{ppaa}) and
(\ref{paap}) for the special choice $p_S = p_L$, $q_S = q_L$ and
$\epsilon = -1$, \cite{BGH99}.

\section{Vanishing of the asymmetry under the Furry hypothesis}
\label{furry hyp}

For $\zeta = 1$ one gets complete decoherence (the Furry hypothesis)
and the interference terms characterizing 2-particle QM correlations
disappear. If furthermore the basis is given by $| f_\pm \rangle$ and the
amplitudes are CP-invariant, the asymmetry vanishes for both $\epsilon
= +1$ and $\epsilon = -1$, \cite{BGH99} (see Eqs.(\ref{A+}) and
(\ref{A-})). One can ask \cite{BGH99}: \textit{What is the basis for
which the asymmetry vanishes under the Furry hypothesis? Is it
necessarily the $| f_\pm \rangle$ basis?} We shall try to answer this
question assuming CP invariance of the amplitudes ($v=V=0$), but not
the WWA. The case of general amplitudes (without CP invariance) can
easily be seen to be pathological. Taking $\epsilon = -1$, one
basically wants $N_1 = 0$, which can be written as
\begin{equation}\label{N=0}
G(t')F(t) + G(t)F(t') = 0
\end{equation}
with the help of Eq.(\ref{N}), using
\begin{equation}
G(t) = |A_2|^2 - |\bar A_2|^2 \,, \quad F(t) = |\bar A_1|^2 - |A_1|^2
\,.
\end{equation}
Thus one wants
\begin{eqnarray}
                 && G = 0 \quad \mbox{if} \quad  F \neq 0 \label{GF} \\
 \mbox{or} \quad && F = 0 \quad \mbox{if} \quad  G \neq 0 \label{FG} \\
 \mbox{or} \quad && F = G = 0 \,.                         \label{FG0}
\end{eqnarray}
To study the consequences, we note that Eq.(\ref{GF}) means
\begin{equation}\label{GF=0}
(|a|^2-|b|^2)(|p_2|^2-|q_2|^2) + 
4\, \mbox{Im}\, (a^*b)\, \mbox{Im}\, (p_2^* q_2) = 0 \,.
\end{equation}
Because of the unmeasurable relative phase between $| f \rangle$ and
$| \bar f \rangle$, the parameter $\mbox{Im}\, (p_2^* q_2)$ can be
chosen arbitrarily. Dropping it, Eq.(\ref{GF=0}) gives
\begin{equation}\label{22}
|p_2| = |q_2| \,,
\end{equation}
because $|a|$ and $|b|$ are in general different. Therefore, with
$\mbox{Im}\, (p_2^* q_2)=0$, we have
\begin{equation}\label{f2+-}
| f_2 \rangle \to | f_+ \rangle \quad \mbox{or} \quad | f_- \rangle
  \,.
\end{equation}
One could have taken $\mbox{Im}\, (p_2^* q_2)=0$ along with
Eq.(\ref{22}) to follow directly from Eq.(\ref{GF=0}) because 
$\mbox{Im}\, (a^*b)$ and $(|a|^2-|b|^2)$ are in general independent
functions of $t$. The argument involving the relative phase becomes
important for the possibility (\ref{FG0}).
Thus, the possibility (\ref{GF}) allows $| f_1 \rangle$ to be
arbitrary, but $| f_2 \rangle$ reduces to a CP eigenstate (see
Eq.(\ref{f2+-})).  One can similarly show that the possibility
(\ref{FG}) allows $| f_2 \rangle$ to be arbitrary, but $| f_1 \rangle$
reduces to a CP eigenstate. The possibility (\ref{FG0}) would mean that
$| f_1 \rangle \to | f_\pm \rangle$ and
$| f_2 \rangle \to | f_\mp \rangle$, thus both basis states become CP
eigenstates. This possibility has been considered in Ref.~\cite{BGH99} in an
explicit calculation using the full WWA and assuming $| f_{1,2} \rangle$ to
form an orthonormal system. Then one easily sees that the three possibilities
(\ref{GF}), (\ref{FG}) and (\ref{FG0}) effectively merge into one, 
and the two basis states reduce
to $| f_\pm \rangle$. The present derivation shows that this last possibility
is allowed but not the most general one: it is sufficient to have only one of
the basis states becoming a CP eigenstate and still achieve $N_1 = 0$.

\section{CP violation due to the introduction of the decoherence parameter}
\label{CP violation}

Starting with the initial state (\ref{state}), let us consider the production
of two final states which are CP conjugates of each other with the respective
QM probabilities $P$ and $\bar P$. Then we can make the decomposition
\begin{equation}
P = P_\mathrm{Furry} + P_\mathrm{int} \quad \mbox{and} \quad
\bar P = \bar P_\mathrm{Furry} + \bar P_\mathrm{int} \,.
\end{equation}
The introduction of $\zeta$ replaces $P$ and $\bar P$ by $Q$ and $\bar Q$,
respectively: 
\begin{equation}
Q = P - \zeta P_\mathrm{int} \quad \mbox{and} \quad
\bar Q = \bar P - \zeta \bar P_\mathrm{int} 
\end{equation}
giving
\begin{equation}
Q - \bar Q = 
(P - \bar P) - \zeta (P_\mathrm{int} - \bar P_\mathrm{int}) \,.
\end{equation}
If under certain conditions $P - \bar P$ vanishes 
(e.g., Eq.(\ref{paap})), we would get
\begin{equation}\label{QQ}
Q - \bar Q = -\zeta ( P_\mathrm{int} - \bar P_\mathrm{int} ) \,.
\end{equation}
This difference would not vanish, because $P = \bar P$ does not
in general imply $P_\mathrm{int} = \bar P_\mathrm{int}$. Eq.(\ref{QQ})
is a simple way of seeing the development of CP non-invariance proportional to
$\zeta$, for any $t'$, $t$. In detailed explicit calculations with the WWA for
$\epsilon = -1$, this has been noticed in Refs.~\cite{BGH99,BG98}. The present
derivation shows the result $Q - \bar Q \neq 0$ due to $\zeta \neq 0$ to be
general. 

An interesting case arises for the like-flavoured final states
$| f \otimes f \rangle$ and $| \bar f \otimes \bar f \rangle$ for 
$\epsilon = -1$ with $t' = t$. Here, both $P$ and $\bar P$ vanish as a simple
consequence of quantum mechanics. On the other hand, 
$Q - \bar Q = -\zeta ( P_\mathrm{int} - \bar P _\mathrm{int} )$
need not vanish due to $\zeta$. This is merely a violation of quantum
mechanics. Though one
may emphasize $Q - \bar Q \neq 0$ as being CP violation, the important point
is that due to $\zeta$ both $Q$ and $\bar Q$ are non-zero in general because
$P_\mathrm{int}$ and $\bar P_\mathrm{int}$ need not vanish for 
$t'=t$. This ``CP violation'' ($Q \neq \bar Q$) 
has recently been noted for $t'=t$ in an explicit
calculation within the WWA and with the further assumption 
$p_S = q_L = q_S = q_L$ for orthonormal basis vectors 
$| f_{1,2} \rangle$ \cite{BGH99}.
The present derivation shows the simplicity and generality of $Q \neq 0$ and
$\bar Q \neq 0$ in this context for $\epsilon = -1$ and $t'=t$.

\section{Summary}

In this paper we have derived some results concerning 2-particle
correlations in the time evolution of the states 
$| f \otimes \bar f + \epsilon \bar f \otimes f \rangle$, 
($\epsilon = \pm 1$), of neutral flavoured mesons and antimesons. We
have put emphasis on deriving our results independently of the
Weisskopf -- Wigner approximation which was used only for comparison
with the general results. We have studied the role of the decoherence
parameter $\zeta$, introduced as a particular way of parameterizing
deviations from quantum mechanics by interpolation between QM 
interference ($\zeta =
0$) and complete loss of interference ($\zeta = 1$, Furry's hypothesis).
Since the QM interference term depends on the basis chosen in the 
$(f, \bar f)$ space, the procedure of introducing the decoherence parameter
is a basis-dependent prodecure; we have tried to elucidate this point 
without having recourse to the specific time evolution of
the WWA. In particular, we have considered the asymmetry (\ref{asymm}) of
unlike-flavoured and like-flavoured events as a 
function of the basis chosen and
of the decoherence parameter. We have shown that several features,
derived previously in the framework of the WWA, persist in general.
An example is the proportionality of the asymmetry to
$(1-\zeta)$, valid for both $\epsilon = +1$ and $-1$, 
if we choose the basis of CP eigenstates and neglect CP
violation in the amplitudes. 
We have derived some other related results more generally
than in the literature
in Sections \ref{purely}--\ref{CP violation}.


\begin{thebibliography}{10}

\bibitem{BGH99}
R.A. Bertlmann, W. Grimus and B.C. Hiesmayr,
Phys. Rev. D \textbf{60}, 114032 (1999).

\bibitem{BG97}
R.A. Bertlmann and W. Grimus,
Phys. Lett. B \textbf{392}, 426 (1997).

\bibitem{dass}
G.V. Dass and K.V.L. Sarma,
Eur. Phys. J. C \textbf{5}, 283 (1998).

\bibitem{BG98}
R.A. Bertlmann and W. Grimus,
Phys. Rev. D \textbf{58}, 034014 (1998).

\bibitem{eberhard}
P.H. Eberhard, in \textit{The Second Da$\Phi$ne Physics Handbook}, edited by L.
Maiani, G. Pancheri and N. Paver (SIS--Pubblicazioni dei Laboratori di
Frascati, Italy, 1995), Vol.~I, p. 99.

\bibitem{WWA}
V. Weisskopf and E. Wigner,
Z. Phys. \textbf{63}, 54 (1930); \textit{ibid.} \textbf{65}, 18 (1930).

\bibitem{CPLEAR98}
CPLEAR Collaboration, A. Apostolakis \textit{et al.},
Phys. Lett. B \textbf{422}, 339 (1998).

\bibitem{furry}
W.H. Furry,
Phys. Rev. \textbf{49}, 393 (1936).

\bibitem{khalfin}
L.A. Khalfin,
Found. Phys. \textbf{27}, 1549 (1997).

\bibitem{dass99}
G.V. Dass,
Phys. Rev. D \textbf{60}, 017501 (1999).

\bibitem{phase}
P.H. Eberhard,
Phys. Rev. Lett. \textbf{16}, 150 (1966).

\bibitem{kabir}
P.K. Kabir,
in \textit{Springer Tracts of Modern Physics}, Vol.~\textbf{52},
edited by G. H\"ohler
(Springer-Verlag Berlin, Heidelberg, New York, 1970), p. 91.

\bibitem{dass92}
G.V. Dass,
Phys. Rev. D \textbf{45}, 980 (1992); \textbf{49}, 1672(E) (1994).

\end{thebibliography}
\end{document}